\documentclass[a4paper,11pt]{article}
\usepackage{jheppub} 
\usepackage{xcolor}
\usepackage{multirow}
\usepackage{subcaption}
\usepackage{float}
\newcommand{\K}{{\cal{K}}}
\newcommand{\C}{{\cal{C}}}

\title{\boldmath Hydrogen Atom in a Fuzzy Spherical Cavity}

\author[a]{M. Hrmo}
\author[a,b]{S. Kov\'{a}\v{c}ik\thanks{Corresponding author.}}
\author[a]{P. Rusn\'{a}k}
\author[a]{J. Tekel}

\affiliation[a]{Department of Theoretical Physics, Faculty of Mathematics, Physics and Informatics,\\
Comenius University in Bratislava,\\
Mlynsk\'a dolina, 842 48, Bratislava, Slovakia}
\affiliation[b]{Department of Theoretical Physics and Astrophysics, Faculty of Science, Masaryk University,\\
Brno, Czech Republic}

\emailAdd{samuel.kovacik@fmph.uniba.sk}

\abstract{The fuzzy onion model formed by connecting a set of concentric fuzzy spheres of increasing radius is motivated by studies of quantum space but can also be used to study standard physics. The main feature of the model is that functions in three-dimensional space -- like scalar fields or wavefunctions -- are expressed in terms of Hermitian matrices of a certain structure. Relevant equations are then matrix equations, and some problems, such as searching for the energy spectrum for fixed quantum numbers $(l,m)$, can be expressed as an eigenvalue problem. We show how this simple approach can reproduce the results of other studies analyzing the hydrogen atom in a spherical cavity. We also test the effect of the short-distance quantum structure of the space on these solutions --- not looking for the phenomenological consequences, as the scale of quantum space is many orders below the order of the Bohr radius, but to understand the effect of quantum space in general. We observe a set of solutions without a classical counterpart which have been suggested also in a former theoretical study.}

\begin{document}
\maketitle
\flushbottom

\section{Introduction}
\label{sec:intro}

The hydrogen atom is one of the most important objects in physics. Thanks to its abundance, it has been thoroughly studied experimentally, analytically, and numerically. In quantum mechanics, it serves an important role as black holes do in general relativity --- it is a rare real example of an exactly solvable problem. In this paper, we present a solution to the problem of hydrogen atoms confined to a spherical cavity. We came across this solution when analysing the behaviour of quantum space, a mathematical formulation that seeks to describe the physics of the Planck length scale. 

The Planck scale is obtained by combining the fundamental physical constants, $G, \hbar, c$, into objects with dimensions of length, mass, energy, etc. From their values, one can infer that the space should not be continuous as is usually assumed in physical models but obtains a quantum structure on the scale of $\mbox{l}_{\scriptsize{Pl}} \approx 10^{-35}m$. Models of quantum space have been considered in the earliest days of quantum mechanics, perhaps as a regularisation tool, but were eclipsed by methods of renormalization. They were brought back to the spotlight when the community started discussing models of quantum gravity, for which scales given by all three aforementioned constants are relevant. Moreover, they appear in certain aspects of the string theory and matrix models of M-theory. This theory has not yet been constructed; there are various promising ideas and approaches \cite{doplicher, Chamseddine:1996zu, Ambjorn:2004qm, Szabo:2001kg, Steinacker:2011ix,Seiberg:1999vs,Berenstein:2015pxa,Steinacker:2022kji} and many of them share a common feature of quantum space \cite{Hossenfelder:2012jw, Nicolini:2022rlz}. 
A quantum space is not a completely well-defined term, as different lines of research perceive it differently. A common way of thinking about it is that it is a space with limited spatial resolution. Any experiment trying to distinguish points below a certain scale, defined by a parameter of length here denoted $\lambda$, which is reasonably assumed to be of the Planck length scale, will fail \cite{Hossenfelder:2012jw}. For example, a particle with a wavelength short enough to distinguish between such points would form a black hole, hiding any desired information. 

Again, there are many approaches for describing a quantum space, some from the phenomenological point of view and some from the mathematical one. The phenomenological aspects are becoming increasingly relevant as technological progress is pushing the limits of indirect observations closer and closer to the Planck length, mostly in the context of astrophysical observations \cite{Amelino-Camelia:2008aez, Amelino-Camelia:1997ieq, Burns:2023oxn, Schupp:2009pt}. 

An important model of quantum space is the fuzzy sphere, which is obtained by putting a cut-off on the spectrum of angular momenta and describing the fields using Hermitian matrices. The size of the matrix is inversely proportional to the spatial resolution --- the larger the matrix size, the higher angular momenta are allowed, and one is able to approximate a positional delta function more and more accurately. The fuzzy sphere model has been studied from various points of view \cite{Hoppe,Madore:1991bw} since it is among the simplest and most instructive models of quantum space, despite the fact that some constructional issues remain unresolved.

Recently, we have formulated a three-dimensional space that is formed as a sequence of concentric fuzzy spheres of increasing radius \cite{Kovacik:2023zab}, built along lines similar to other works \cite{seik, pgk15, pg13, Kupriyanov:2012nb,  todorovic1,todorovic2, sch1, sch2, sch3, sch4,patricia,patricia2,Tekel:2015uza}. The main goal was not to obtain a precise mathematical description of quantum space originating from a fundamental theory, but to obtain a model that carries both the essential features of quantum space and is reasonably amenable to studies of interesting problems --- field theories, classical mechanics, waves on neutron stars, gravitational collapses and models of microscopic black holes. This wide range of areas of applicability stems from the fact that one can change the value of $\lambda$ to a different value, and then it can describe a scale of the granularity of the given material rather than the Planck scale \cite{granular}. 

The fuzzy onion model has various possible applications, and it has been tested using the simplest example of a hydrogen atom problem. Later, we noticed that the formulation of the problem allowed for a simple code to solve the hydrogen problem in a spherical cavity, which is natural for our description of the model since it is constructed by adding spherical layers together. In this paper, we explain the construction of the problem and compare our results with high-precision results obtained mostly in the chemical literature \cite{aquino2007highly}. Therefore, we also express the model in a form that might be useful for the study of other physical-chemical problems. We also test our model in the regime of strong quantum-space effects, perhaps to obtain some insight into the behaviour of matter in the earliest stages of the universe. 

\section{A brief overview of the fuzzy onion model}
\label{sec:fuzzy onion}
Let us begin with a short summary of the fuzzy sphere construction before we move to the fuzzy onion. Fields on the standard sphere, or angular-dependent parts of three-dimensional wave functions, can be expanded in terms of the spherical harmonics, which are eigenfunctions of the angular momentum operators:
\begin{eqnarray} \nonumber
 f(\theta,\varphi) &=& \sum_{lm} c_{lm} Y_{lm} (\theta,\varphi), \\ \label{eq1}
 \hat{{\cal{L}}} \ Y_{lm} &=& \hat{L}_i\ \hat{L}_i \ Y_{lm} = l (l+1) Y_{lm} , \\ \nonumber
 \hat{L}_3 \ Y_{lm}&=& m\ Y_{lm} ,
\end{eqnarray}
where differential operators act via the commutator, $\hat{L}_i Y = [L_i, Y]$. These form an infinite-dimensional representation of the $su(2)$  algebra. There are, however, finite-dimensional representations; the smallest nontrivial one expressed in terms of Pauli matrices might be the most well-known one. For any matrix size $N$, one can find three matrices $L_i$ and $N^2$ matrices $Y_{lm}$ that satisfy relations \eqref{eq1}; details can be found in the appendix \ref{appA}. The matrices $Y_{lm}$ form a basis for Hermitian matrices of the given size, which means any Hermitian matrix can be expanded as
\begin{eqnarray}\label{mapping}
    \Phi = \sum \limits_{l,m} c_{lm}Y_{lm},
\end{eqnarray}
where $l=0,1,\ldots,N-1$ and $m = -l,\ldots, l$. This means that such matrices can be mapped onto functions of angular coordinates $\Phi \rightarrow \phi(\theta,\varphi)$ just by using the same expansion coefficients. The map works both ways as long as we use only functions with maximal momentum in the expansion equal to $N-1$; such functions cannot have arbitrarily small support --- this is what is meant by the limited spatial resolution. These functions can be mapped to Hermitian matrices that form a noncommutative algebra closed under multiplication with a defined set of rotation generators. Other relevant objects can be defined as well; for example, taking the integral of $\phi$ corresponds to taking the trace of $\Phi$. We refer the reader to \cite{Kovacik:2023zab} for a more detailed discussion and move to a brief description of the fuzzy onion model. 

The angular momenta in the expression \eqref{eq1} are dimensionless, a noncommutative algebra defined by $[x_i, x_i] = 2i \lambda \varepsilon_{ijk}x_k$ requires a new parameter with the dimension of length, which is denoted $\lambda$ and defined as $x_i = 2\lambda L_i$. From this, one can derive a relation between the matrix size and radius of the fuzzy sphere: $R =\lambda \sqrt{N^2-1} \approx \lambda N$; here, we take $R = \lambda N$. By keeping $\lambda$ fixed, larger matrices correspond to larger spheres. That is, a sphere of radius $\lambda$ can be described by a matrix of size one, a sphere of radius $2\lambda$ by a $2 \times 2$ matrix and so on. That means that the set of matrices of increasing size, $\left(\Phi^{(1)}, \Phi^{(2)}, \ldots,\Phi^{(M)} \right)$, where the superscript denotes the matrix size, describe a set of $M$ disjoint concentric fuzzy spheres. This set can be placed into a larger matrix $\Psi$ defined as
\begin{equation} \label{psi}
 \Psi = \begin{pmatrix}
\Phi^{(1)} & & &\\
 & \Phi^{(2)} & &\\
 & & \ddots &\\
 & & & \Phi^{(M)}
\end{pmatrix}.
\end{equation}
As the integration on a single fuzzy sphere is replaced by a trace, the same can be done here. There is an important detail --- since $\mbox{Tr}(\ldots)$ corresponds to $\int (\ldots) r \ d \Omega$, one needs to add a certain factor to obtain the proper measure:
\begin{eqnarray} \label{norm}
    \int d^3 \psi \leftrightarrow \mbox{Tr} \left( 4 \pi r \Psi \right),
\end{eqnarray}
where $r$ is a diagonal matrix with entries $\lambda, 2\lambda,2\lambda, 3 \lambda,3 \lambda,3 \lambda, 4 \lambda, \ldots$ . As we can define the angular part of the Laplace operator, that is $\Delta_L = - \frac{{\cal{L}}}{r^2}$, on each of the spheres individually, it can be defined on the entire matrix $\Psi$ as well:
\begin{equation} \label{L}
 {\cal{L}} \Psi = \begin{pmatrix}
 \hat{{\cal{L}}}^{(1)}\Phi^{(1)} & & &\\
 & \hat{{\cal{L}}}^{(2)}\Phi^{(2)} & &\\
 & & \ddots &\\
 & & & \hat{{\cal{L}}}^{(M)}\Phi^{(M)}
\end{pmatrix}.
\end{equation}
Defining the radial part of the kinetic term requires a certain nuance, as one cannot directly subtract matrices of different sizes, as is needed when computing the radial derivative. This issue can be solved by defining two operators, ${\cal{U}}$ and ${\cal{D}}$ that decrease and increase the matrix size; details can be found in the appendix \ref{appB}. Using those, we can define
\begin{equation} \label{dr}
 \partial_r \Phi^{(N)} = \frac{{\cal{D}}\phi^{(N+1)} - {{\cal{U}}}\phi^{(N-1)} }{2\lambda},
\end{equation}
and 
\begin{equation} \label{ddr}
 \partial^2_r \Phi^{(N)} = \frac{{\cal{D}}\phi^{(N+1)} -2\phi^{(N)} + {{\cal{U}}}\phi^{(N-1)} }{\lambda^2},
\end{equation}
Then, the full Laplace operator acting on the fuzzy onion states can be defined as 
\begin{equation}
 \K \Psi = \left(\frac{1}{r^2} \frac{\partial}{\partial_r} \left( r^2 \frac{\partial}{\partial r} \right) - \frac{{\cal{L}}}{ r^2} \right) \Psi,
\end{equation}
which serves as the kinetic energy operator in the Hamiltonian. We are now ready to analyze the problem of the Hydrogen atom in the cavity in this setting.

\section{The hydrogen atom in the spherical cavity}
\label{sec:Schrodinger}

In this section, we describe our calculation and illustrate the most important findings. More detailed results are discussed in the appendix \ref{appC}.

\subsection{Construction and standard states}
In ordinary quantum mechanics, the hydrogen atom in a cavity is described by the Hamiltonian
\begin{eqnarray}
    H = -\frac{1}{2 m}\K - \frac{q^2}{r} + V(r),
\end{eqnarray}
where $V(r)=0$ for $r<r_0$ and ${V(r)=\infty}$ otherwise and $r_0$ is the size of a spherical cavity. Here, $q$ and $m$ are the charge and the mass of the electron, respectively, and will be set to $1$ from now on; we also use units where $\hbar=1$, which also sets the Bohr radius and the Hartree energy to $1$. 

In our construction, the cavity size is given by $r_0=M \lambda$, where $M$ is the size of the matrix describing the outermost layer. This can be viewed as either a fundamental quantum feature of the space or as a regularisation tool for the continuous space limit obtained by taking $M\to\infty$ and $\lambda\to 0$ while keeping $r_0$ fixed.

When solving this problem in standard quantum mechanics, one can exploit the rotational symmetry and split the solution into the form $\psi(r,\theta,\phi) = R(r) Y(\theta,\phi)$. Something similar can be done in the case of the fuzzy onion model. When fixed values of $(l,m)$ are chosen, then each layer is represented by a single coefficient $c^{(i)}_{lm}$ and together they form a vector, dropping the trivial terms:
\begin{eqnarray}
 {\cal C}_{lm} = \left(c^{(l+1)}_{lm} ,\ldots,c^{(M)}_{lm}\right),
\end{eqnarray}
and in this vector representation, we can write the Schrödinger equation in the form
\begin{align} \label{Schrodinger}
    \mathbf{H} {\cal C}_{lm} =E {\cal C}_{lm},
\end{align}
where
\begin{align}  
    \mathbf H=-\frac{1}{2} \left(\K _R-l(l+1)\mathbf r^{-2}\right)-\mathbf r^{-1} \label{num:ham1},
\end{align}
we denote the operators acting on vectors $\mathcal C$ rather than matrices $\Phi$ in boldface and
$\K _R$ is specified in the appendix \ref{appB}.
Putting all the terms together, one obtains a single large matrix $\mathbf{H}$, its eigenvalues are the energy eigenvalues of the Hamiltonian, and the eigenvectors form the radial part of the solution.

We observe that eigenvalues generally split into two classes: one that corresponds to the bounded states and one that corresponds to scattering states, or to be more precise --- the states that would correspond to those in the infinity-cavity limit; we separate them by the sign of their energy. We have evaluated the eigenvalues for various combinations of $r_0, M, n, l$ and observed a considerable agreement with the reference \cite{aquino2007highly}; the comparison is shown in tables in the section \ref{Numerical results}. The eigenvectors corresponding to the Hamiltonian eigenvalues are the radial parts of the energy eigenstates, and we can use mapping \eqref{mapping} to reconstruct the radial probability distribution for the electron. Two examples are shown in Figures \ref{fig1} and \ref{fig2}. We can see that our matrix description of the model can be used to efficiently reproduce the results from ordinary quantum mechanics.

We can compute the mean value of observables defined as
\begin{eqnarray}
    \left\langle \mathcal{O} \right\rangle = \mbox{Tr} \left( 4 \pi r  \mathcal{O}\right).
\end{eqnarray}
Note that the observables are, like other relevant objects, matrices. As an example, we compute $\left\langle r \right\rangle$; the results are shown in tables \ref{r1} and \ref{r2}. We show the results for two different values of $M$, which, when comparing the same values of $r_0$, relate to two different values of $\lambda$. It is clear and consistent that by increasing the value of $\lambda$, the particle is pushed outward from the centre. This suggests that the quantumness of space creates a form of repulsion. This insight can be useful for models of matter singularities in a quantum space. 

Let us now consider the continuous space limit. If we keep $r_0$ fixed and increase $M$, the scale of noncommutative $\lambda$ vanishes, and we reproduce the model of the hydrogen atom in the spherical cavity of radius $r_0$ in ordinary quantum mechanics. One can also do a double limit and send $M \rightarrow \infty$, which reproduces the ordinary hydrogen atom spectrum. When studying the confined hydrogen, three regimes are of interest: a strongly compressed particle with $r_0<a_0$, where $a_0$ is the Bohr radius, equal to unity in our units. The second regime is a mildly compressed atom with $r_0 \approx a_0$ and then a barely confined atom with $r_0 \gg a_0$. These, all investigated in the commutative limit, are interesting from a practical point of view in a chemical context \cite{Reyes2024, Scherbinin1998, AlHashimi2015,Reyes2024a, AlHashimi2012,Kamel2016, Goldman1992, Aquino2009, Laughlin2002, White2015, Ping2019, Chaudhuri2021, Fischer2023, aquino2007highly}. Basically, the hydrogen atom placed in a cavity is a useful model to study changes in reactions confined to a small reaction site or under high pressure. Therefore, it might be relevant to analyze more complex chemical compounds in this setting; we leave this for future study.  

\subsection{Numerical results}
\label{Numerical results}
In this section, we gather the results of numerical simulations performed on a personal computer. The reference energies denoted $\infty$ as ${M\rightarrow \infty}$ correspond to the limit of ordinary space and are taken from \cite{aquino2007highly}. The results provided are eigenvalues of the matrix defined in \eqref{Schrodinger}, shown for different numbers of layers. The energies are in hartrees and the distances in bohrs. 

The large matrix values shown in Table \ref{rlargeN} were obtained in 10 hours on a personal computer, and the code was written in C\texttt{++}. It is not clear whether deviations are due to numerical errors in our computation or the approximations in the reference used. We can observe that a reasonably small matrix can be used to reproduce the value of the lowest energy level accurately. 

\begin{table}[h!] 
\centering
    \centering
    \begin{tabular}{|c||c|c|c|c|c|c|}
        \hline
        $ M$   & $10^1$      & $10^2$      & $10^3$ & $10^4$  &  $10^5$ & reference \\ \hline
        $|E_1|$ & $0.41421243$ & $ 0.49875557$ & $0.49998677$ & $0.49999919$ & $0.49997571$ & $0.49999926$\\ \hline
    \end{tabular}
    \caption{Values of the ground state energy for $r_0=10$ and different number of layers. The largest value took approximately $10$ hours of computational time on a personal computer. The reference is taken from \cite{aquino2007highly}.}
    \label{rlargeN}
\end{table}

\begin{table}[h!] 
\centering
\begin{minipage}{.45\linewidth} 
    \centering
    \begin{tabular}{|c||c|c|c|}
        \hline
        $r_0$   & 1S      & 2S      & 3S      \\ \hline
        0.5 & 0.24251 & 0.25061 & 0.25083 \\ \hline
        1   & 0.46836 & 0.50337 & 0.50361 \\ \hline
        4   & 1.34177 & 2.14646 & 2.08655 \\ \hline
    \end{tabular}
    \caption{Mean value of $\left\langle r \right\rangle$ (in bohrs) for $M = 10^4$ and $l=0$.}
    \label{r1}
\end{minipage}
\hspace{1cm} 
\begin{minipage}{.45\linewidth}
    \centering
    \begin{tabular}{|c||c|c|c|}
        \hline
        $r_0$  & 1S      & 2S      & 3S      \\ \hline
        0.5 & 0.24484 & 0.25310 & 0.25332 \\ \hline
        1   & 0.47265 & 0.50841 & 0.50864 \\ \hline
        4   & 1.34756 & 2.16978 & 2.10845 \\ \hline
    \end{tabular}
    \caption{Mean value of $\left\langle r \right\rangle$ (in bohrs) for $M = 10^2$ and $l=0$.}
    \label{r2}
\end{minipage}
\end{table}

In tables \ref{r1} and \ref{r2}, the mean value $\left\langle r \right\rangle$ is shown for various orbitals and two different cavity sizes; they are different in the size of the used matrix. A smaller matrix with the same cavity radius corresponds to a coarser underlying structure; that is, the relative size of quantum cells is larger, and the effects of quantum space are more prominent. We can observe that for small matrix sizes, the mean value of the radial coordinate of the particles increases when they are pushed outward from the centre. 

We are also interested in the regime of strong effects of quantum space. We can again consider three situations with regard to the size of confinement, but now, instead of taking $\lambda \rightarrow 0$, we take $\lambda \rightarrow a_0$. Why is this situation interesting? Firstly, it can give some clues about the behaviour of particles in the earliest stages of the universe, where everything was dense and compressed and also energetic enough to make the effects of quantum space prominent. Secondly, it can provide some insight into gravitational collapse. Of course, in our situation, there is no gravity, but earlier studies \cite{Nicolini:2005vd} suggest that the effects of quantum space prevent the formation of a perfect singularity and provide a form of outward pressure that stabilizes the core, so we can now see whether the quantumness of space would act against the squeezing force of the confinement. 

\begin{figure}[h!]
\centerline{\includegraphics[width=1.1\textwidth]{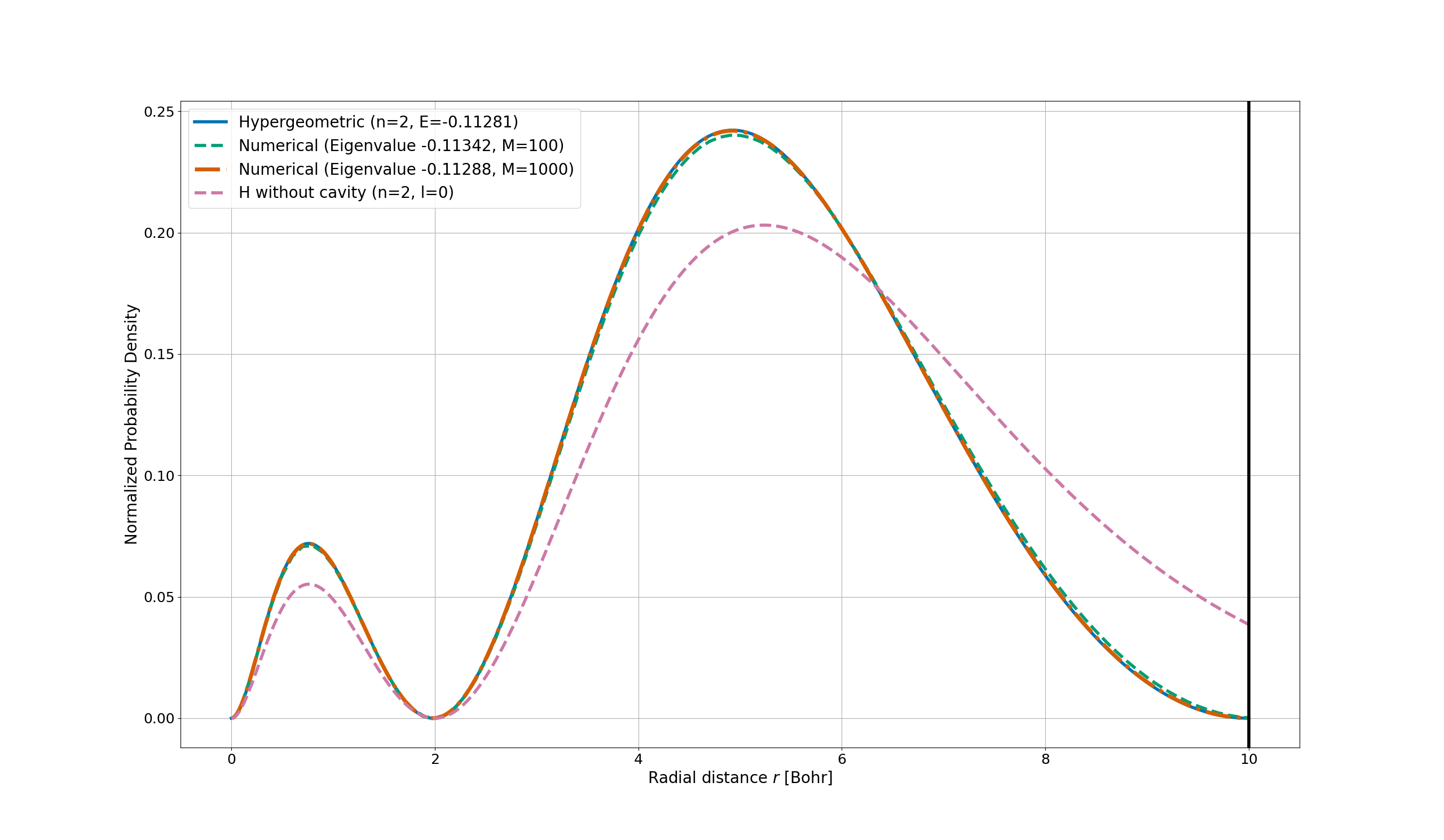}}
    \caption{Comparison of the radial part of our solution using the matrix defined  \eqref{Schrodinger} and the ordinary-space solution for confined hydrogen atom from reference \cite{aquino2007highly} for $r_0 = 10$ and \textbf{2S} orbital. We have also included the unconfined solution to show the effect of the barrier.}
\label{fig1}
\end{figure}

While the eigenvalues of \eqref{Schrodinger} give us the energy spectrum, the corresponding eigenvectors are the radial part of the solution $\Psi$. In Figure \ref{fig1}, we show one particular example. We can see that our solution satisfactorily matches the solution obtained in \cite{aquino2007highly}, and the deviation from the unconfined solution is also prominent. We have made a similar comparison for other orbitals and values of parameters, and the matching holds for as long as we are far away from the strong quantum space regime, that is, as long as $\lambda$ remains sufficiently small.

\begin{figure}[h!]
\centerline{\includegraphics[width=1.1\textwidth]{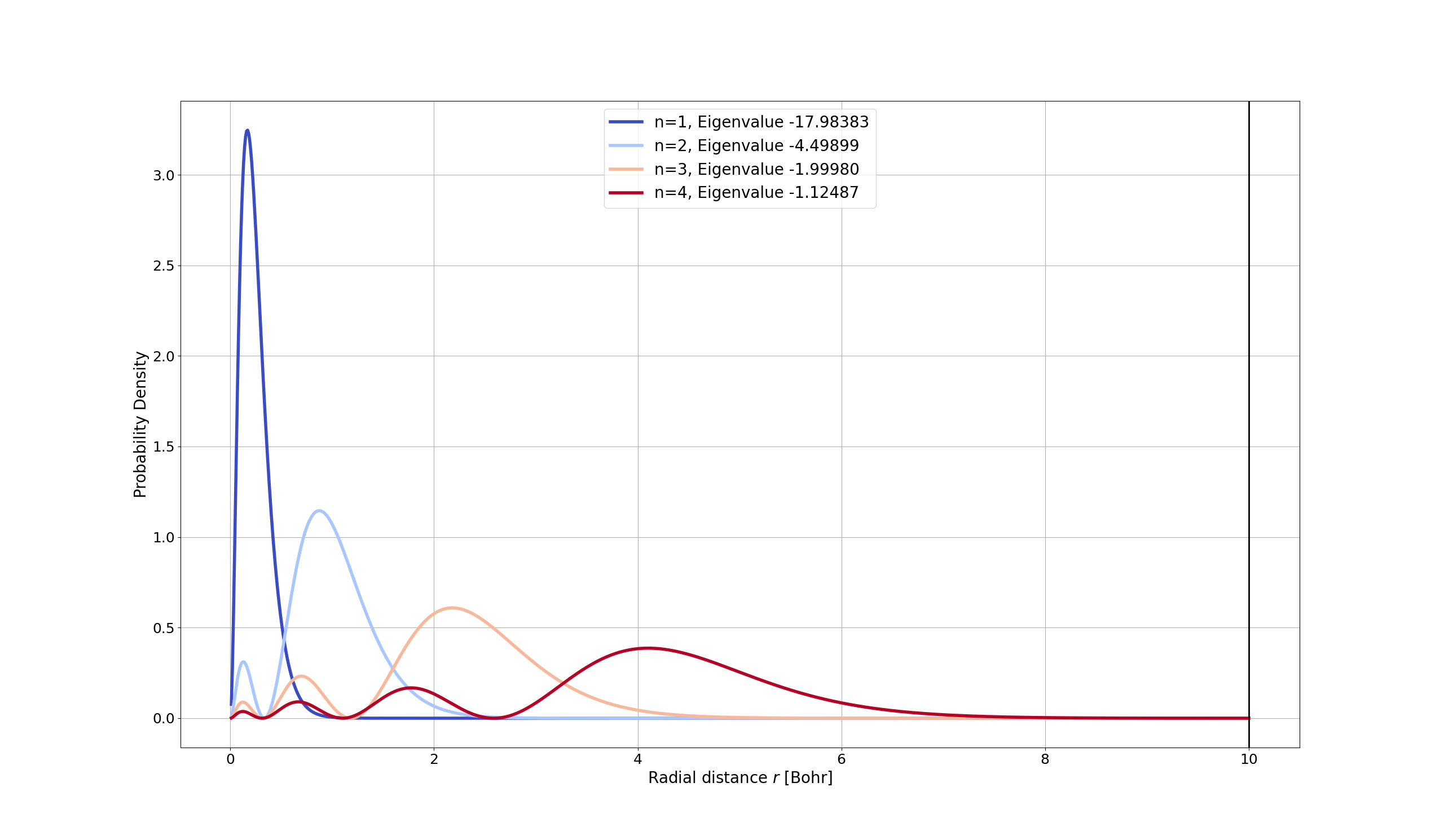}}
    \caption{First four states of Coulomb problem with additional linear term $H_{\scriptsize{add}}={-5r}$ in Hamiltonian. $M = 1000$, $r_0 = 10$ and $l=0$.} 
\label{double_conf}
\end{figure}

An important aspect of our construction is that it can be easily modified to include various terms in the potential --- we have focused on the hydrogen model as on an illustrative example. To demonstrate this, we have added a linear term 
\begin{eqnarray}
    H_{\scriptsize{add}} = - 5 r,
\end{eqnarray}
to the Hamiltonian \eqref{Schrodinger} and found its spectrum and set of solutions, the first four of them are shown in Figure \ref{double_conf}. This potential has been analysed in the context of the study of bound states of quarks using different methods in \cite{Bukor:2022xjc}. This shows that the constructions have the potential to solve precisely many complicated quantum-mechanical models, at least those possessing rotational symmetry. 

\subsection{Mirroring states}
\begin{figure}[h!]
\centerline{\includegraphics[width=1.1\textwidth]{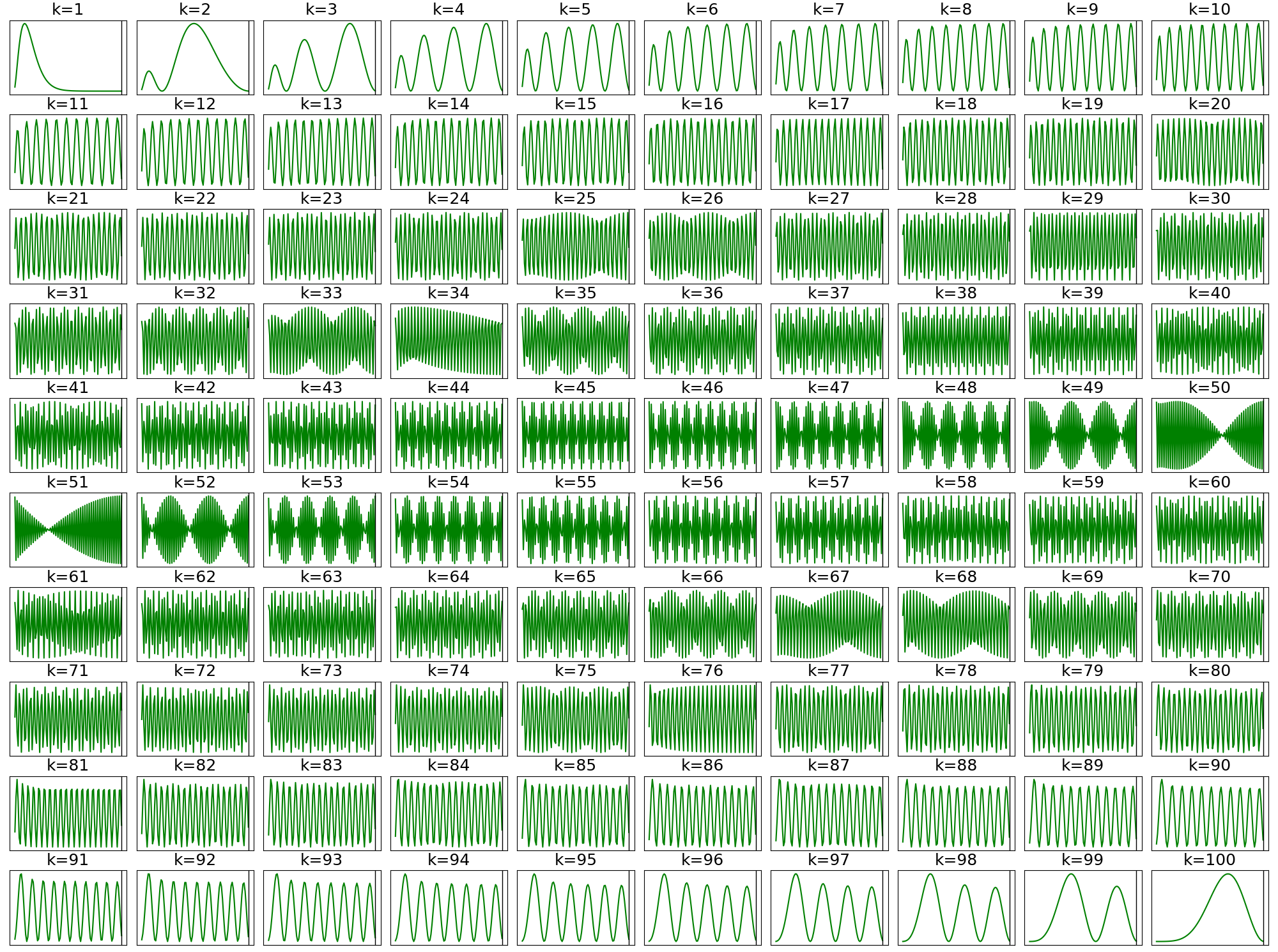}}
    \caption{Radial parts of the probability distribution for all energy eigenstates for the model \eqref{num:ham1} with $r_0=10$ and $M=100$. The same plot is obtained --- in reverse order, starting from the largest eigenvalue, see \eqref{dual energy} --- for the same model but with the repulsive potential. Green lines show numerical solutions of \eqref{Schrodinger}.} 
\label{radial_wavefunctions}
\end{figure}

We have observed a class of solutions that have no counterpart when the space is continuous, that is ${\lambda=0}$. These have been conjectured in a previous study of the hydrogen atom in a noncommutative space \cite{pg13, Gal:2013}. It has been observed that in addition to states with energies $E_n^I$ that reproduce the spectrum of an unconfined hydrogen atom in ordinary space, there is an additional set of bound states for a repulsive potential of the same form. The value of these energies is
\begin{align} \label{dual energy}
    E_n^{II} = \frac{2}{\lambda^2}-E_n^{I}.
\end{align}
In Table \ref{EIEII}, we compare the two energy scales obtained as smallest and largest eigenvalues of \eqref{Schrodinger} with the values predicted by \eqref{dual energy} --- the match is nearly perfect. In Figure \ref{radial_wavefunctions}, we plot the radial part of the wavefunctions for an attractive potential. We obtained essentially identical results for the repulsive potential, taking \eqref{dual energy} into account.

\begin{table}[h!] 
\centering
\resizebox{1\textwidth}{!}{
\begin{tabular}{|c||c|c|c|c|}
\hline
$n$  & $E_n^{II}$             & $E_n^{I}$        & $\frac{2}{\lambda^2} - E_n^{II}$               & difference           \\ \hline
1 & 20000.499986778540915 & -0.499986776755433 & -0.499986778540915 & 0.000000001785482 \\ \hline
2 & 20000.112878189818730 & -0.112878188196622 & -0.112878189818730 & 0.000000001622108 \\ \hline
3 & 19999.909082941998349 & 0.090917058742129  & 0.090917058001651  & 0.000000000740478 \\ \hline
4 & 19999.596027779884025 & 0.403972221239958  & 0.403972220115975  & 0.000000001123983 \\ \hline
\end{tabular}}
\caption{$E^{II}$ are the largest eigenvalues for repulsive potential, $E^{I}$ are the smallest eigenvalues for the attractive potential. The next column shows the theoretical prediction, and the last one compares this with the obtained numerical values. Energies are in Hartree.}
\label{EIEII}
\end{table}

A similar mirror has been observed for other choices of the potential $V(r)$. Here, we sketch the proof and underlying assumptions for the existence of mirroring spectra. That is, for a set of solutions of the eigenproblem with $V(r) \rightarrow -V(r)$ and energy spectrum \eqref{dual energy}, that satisfy:
\begin{eqnarray}
    \left(\textbf{H}_0 + \textbf{V}(r)\right)  {\cal C}&= E_I  \ {\cal C}, \\ \nonumber
      \left(\textbf{H}_0 - \textbf{V}(r)\right)  {\cal C}'&= E_{II} \  {\cal C}'.  
\end{eqnarray}
Since ${\cal C}$ and ${\cal C}'$ are vectors, we can take $ {\cal C}' =  \textbf{P} {\cal C}$, where $\textbf{P}$ is a matrix. The next step is to multiply the first equation by $\textbf{P}$ from the left and add those two equations together, obtaining
\begin{eqnarray}
   \left( \{\textbf{P},\textbf{H}_0\} + [\textbf{V}(r),\textbf{P}]\right) {\cal C} = \left(E_I+E_{II}\right) \textbf{P} {\cal C}.
\end{eqnarray}
From this, we deduce two equations for $\textbf{P}$:
\begin{eqnarray}
  \{\textbf{P},\textbf{H}_0\} &=&  \left(E_I+E_{II}\right) \textbf{P} \\ \nonumber [\textbf{P},\textbf{V}(r)]&=&0, 
\end{eqnarray}
which are solved by a diagonal matrix of the form
\begin{equation}
    \textbf{P} = \mbox{diag }\left(1,-1,1,-1, \dots\right),
\end{equation}
with $\left(E_I+E_{II}\right) = 2 \lambda^{-2}$, yielding \eqref{dual energy}. This matrix has a physical interpretation. The state with maximal oscillation in the fuzzy onion space has the form ${\cal C} \sim \left(1,-1,1,-1, \dots \right)$. This matrix maps the constant vector onto this, and since $\textbf{P}^2 = \textbf{1}$, it maps the maximal oscillation vector into a constant one, a highly-oscillatory state is mapped into a nearly constant and so on; connecting the UV part of on theory into IR part of the one with flipped potential. Of some interest might be the state satisfying $\textbf{P} {\cal C} = {\cal C}$. In \cite{KovacIk:2013vbk}, it has been observed that while the energy spectrum is bounded by $2 \lambda^{-2}$, the Heisenberg uncertainty relation vanishes for a (free) state with the energy of $\lambda^{-2}$; it might be worthwhile to investigate where this self-adjoined state has some special properties in the case with nontrivial potential as well.

\subsection{Number of negative eigenvalues}

\begin{figure}[H]
\centerline{\includegraphics[width=1.1\textwidth]{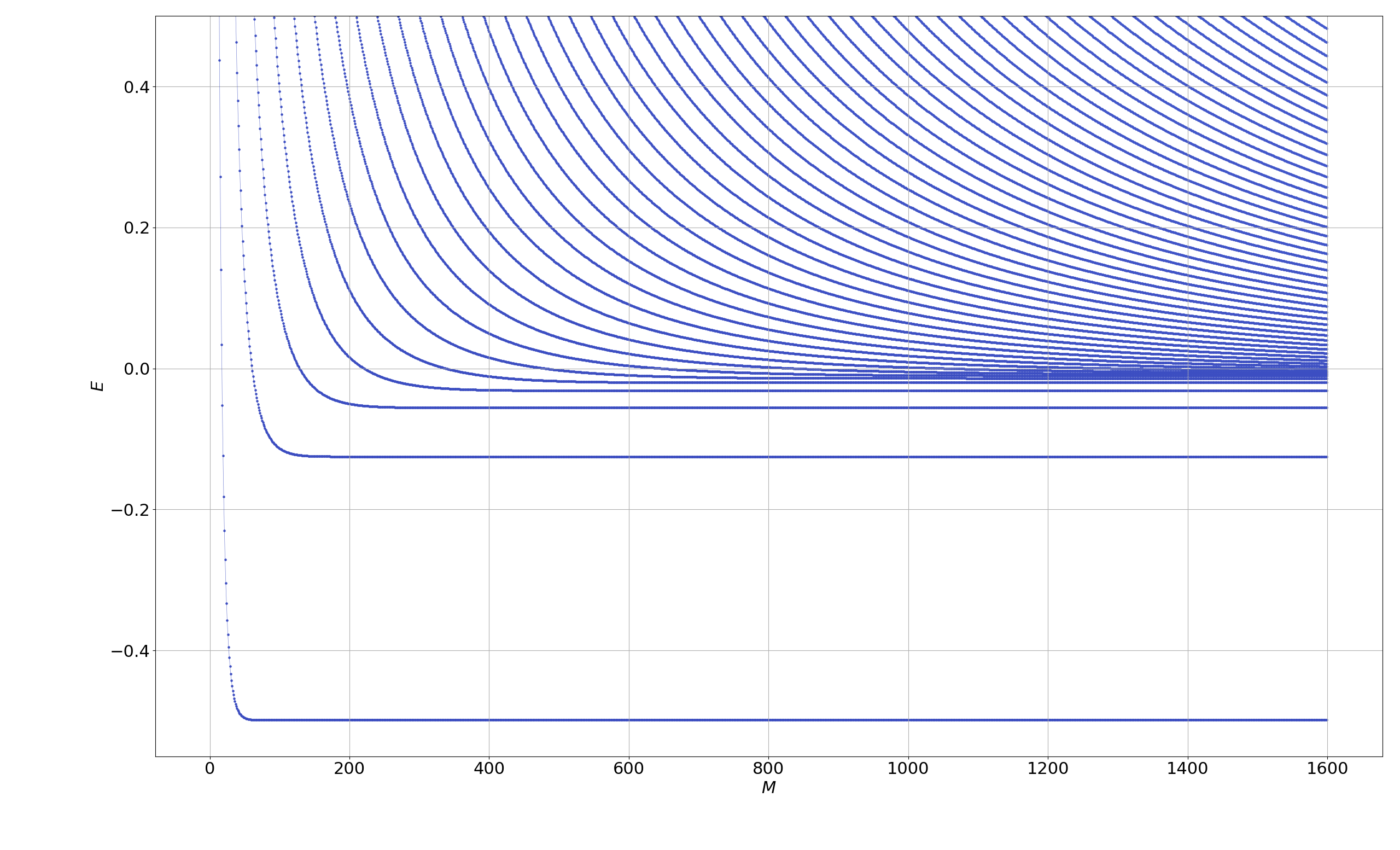}}
    \caption{Spectrum of the hydrogen atom with $\lambda = 0.1$. For better visualisation of the eigenvalue flow, the smallest, second smallest, and so on, eigenvalues have been connected across various values of $M$.} 
\label{eigenflow1}
\end{figure}

\begin{figure}[H]
\centerline{\includegraphics[width=1.1\textwidth]{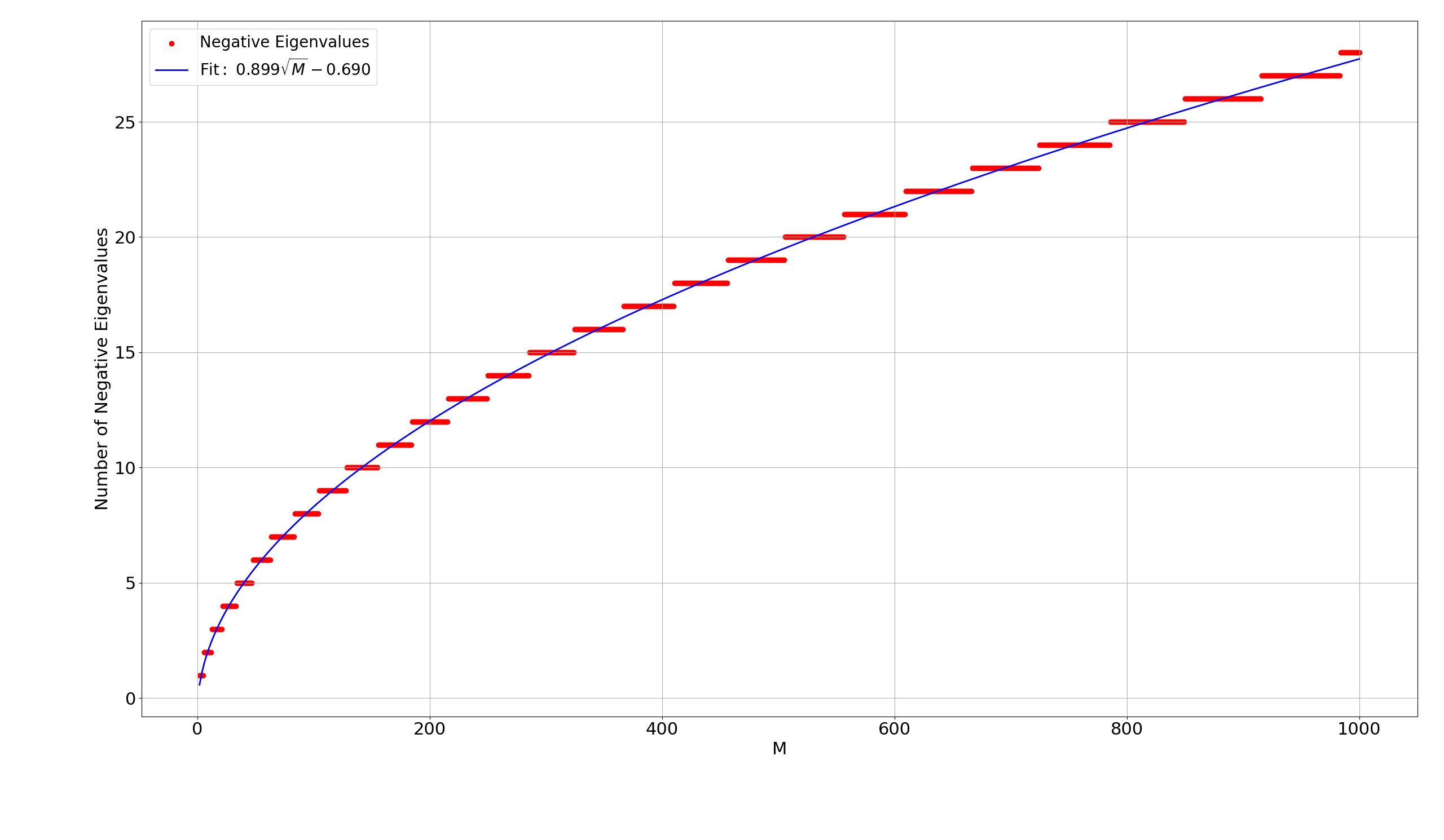}}
    \caption{Number of negative energies of the hydrogen atom with respect to M for $\lambda = 1$. The blue line is fit with the function $a\sqrt{M}+b$} 
\label{numberofnegative01}
\end{figure}

In this paper, we study a finite model that has, as we have argued, a limit in which it approaches ordinary continuum three-dimensional space. Let us take the hydrogen model as an example. In the continuum limit, it has a discrete part of the energy spectrum (bounded states with negative energy) and a continuous part (scattering states with positive energy). For our finite model, the particle is always confined in the cavity, but we can still analyse how, for a fixed set of parameters, the finite number of eigenvalues split between positive and negative values, and how they behave as the continuum limit is approached. To be precise, let us fix $\lambda=0.1$ and see the energy spectrum for different values of $M$ (considering only $l=0$ states). In Figure \ref{eigenflow1}, we can see that the smallest eigenvalue initially was positive and as $M$ was increased, it decreased, became negative and converged to a fixed value. Similar behaviour appeared for the rest of the eigenvalues. To summarise, by increasing the matrix size $M$, the number of eigenvalues grows and their corresponding values gradually decrease and converge to a certain value. We have computed the number of negative eigenvalues for various values of $M$ and the results are shown in Figure \ref{numberofnegative01}. As we can observe, their number grows proportional to $\sqrt{M}$, meaning that in the continuum limit, the ratio of the number of scattering states and bounded states grows indefinitely. There is a physical explanation for the square root growth. The average distance from the origin for a bounded state at energy level $n$ grows as $\langle r \rangle \sim n^2$; at least in the continuum limit with infinite volume, our model is an approximation of this. The radius of the cavity is $R = \lambda M$, and unless the state fits in it, that is $ R \approx \langle r \rangle $, its wavefunction is distorted and energy increased, making it positive. Therefore, the largest negative level $n$, that is the number of bounded states, grows as $n\sim \sqrt{M}$.  

\section{Conclusions and outlook}
\label{sec:conc}

We have demonstrated that a simple model of quantum space can be used, in a certain limit, as a description of ordinary physics, e.g. the hydrogen atom problem. The defining feature of this model is that it is expressed in terms of matrices and can, therefore, easily be analysed on a computer. 

To showcase the behaviour of the model, we have analysed the behaviour of the hydrogen atom confined in a spherical cavity and compared our results to those previously obtained in high-precision studies; the agreement is up to numerical uncertainty. We have set up a Hamiltonian expressed in terms of a matrix; its eigenvalues form the energy spectrum, and the corresponding eigenvectors for the radial dependency. The model can be easily amended to have a different form of potential, making this a rather flexible model.

In the case of the hydrogen atom, it was possible to take only a single value $(l,m)$; however, angular-dependent problems can be analyzed in the defined setting as well; one just needs to use the full form of the matrix $\Psi$ and the Hamiltonian. 

In our case, we sent the constant $\lambda$ describing the quantumness of space to $0$ to recover the physics in ordinary space. However, one can be motivated to choose a nonzero value to describe a space with a granular structure or to capture the essence of a finite-length scale feature of the studied model. 

We have verified the behaviour of the confined particle with respect to the different radii of the cavity and showed how the particle is affected in certain regimes. We have also studied the behaviour in the strong quantum-space regime. The first important observation was that increasing the value of $\lambda$ pushes the particle away from the centre; this means that the quantumness of space manifests as an outward pressure. This is important, for example, to understand the behaviour of regularized singularity in studies of microscopic black holes sensitive to this length scale \cite{Nicolini:2005vd, Kovacik:2021qms}. 

In addition to that, it has been shown that a new form of bounded states should appear for ultra-high energies \cite{pgk15}. We have shown that these states indeed appear and confirm the pattern of mirroring energies. Since the parameter, $\lambda$ can also be understood as describing a length scale that is not of quantum-space origin, but perhaps due to two granular features of the material, it will be interesting to consider whether such mirroring states might be experimentally observable. Also, our plan is to investigate the behaviour of a free particle moving in confinemed region, studying the pressure it produces (in the strong quantum space limit); those studies should provide some insight into high-density regimes of the theory of quantum space, which are needed for the description of the earliest stages or the universe or the behaviour of black hole singularities as described in \cite{pgk15, pg13}. 

\acknowledgments

This research was supported by VEGA 1/0025/23 grant \emph{Matrix models and quantum gravity} and MUNI Award for Science and Humanities funded by the Grant Agency of Masaryk University. The authors also thank the hospitality of the Physikzentrum Bad Honnef, where the work in this paper was initiated. The authors would like to acknowledge the contribution of the COST Action CA23130. We would like to thank Lukáš Konečný for his comments. 

\appendix
\section{Explicit construction of the $su(2)$ representation with Hermitian matrices of rank $N$}\label{appA}
Let us begin with the construction of the angular momentum generators satisfying \begin{equation}
    [L_i, L_j]=i\varepsilon_{ijk}L_k.
\end{equation}
We can choose a basis for which eigenstates of $L_3$ are unit vectors of the form
\begin{equation}
e_{l-m}=(0,\ldots,0,1,0,\ldots,0)    
\end{equation} such that $\hat{L}_3 e_m =m e_m$. Then, we define two matrices that increase this value by $1$, that is $L_{
\pm} e_m \sim e_{m\pm 1}$. Taking into account the normalisation, one obtains
\begin{equation}
    L_3 = \begin{pmatrix}
 m & 0 & 0 & \cdots \\
 0 & m-1 & 0 &\cdots \\
 0 & 0  & m-2 &\cdots  \\
 \vdots & \vdots & \vdots & \ddots
\end{pmatrix}, L_+ = \begin{pmatrix}
 0 & \sqrt{l} & 0& 0& \cdots\\
 0&0 &\sqrt{2 \left(l-1\right)} & 0 & \cdots\\
 \vdots & \vdots & \vdots & & \ddots  
\end{pmatrix},
\end{equation}
and $  L_- = L_+^T$. The off-diagonal coefficients are $\sqrt{\left(l+1  - m\right) m}$. Now one can take 
\begin{equation}
    L_1 = \frac{L_++L_-}{2}, L_2 = \frac{L_+-L_-}{2i}.
\end{equation}
Eigenstates to $\hat{L}_3$ and $\hat{L}^2$ can be found first by finding $Y_{l,0}$ in terms of diagonal matrices and then generating those with nonzero values of $m$ by the actions of $\hat{L}_{\pm}$. For a suitably chosen normalisation, some useful relations hold; see \cite{Tekel:2015uza}.

\section{Radial term on the fuzzy onion}\label{appB}
The action of the radial part of the kinetic term on the full matrix $\Psi$ is
\begin{equation} \label{KR2}
   {\cal{K}}_R \Psi = \sum \limits_{N,l,m}\frac{(N+1) c^{(N+1)}_{lm} + (N-1) c^{(N-1)}_{lm} - 2 N c^{(N)}_{lm}}{N\,\lambda^2} Y^{(N)}_{lm}.
\end{equation}
After restricting only on states with a fixed value of $(l,m)$, we can drop the summation over those indices. The state is now represented by the vector of coefficients 
\begin{align}
\C_{lm}=\left(c^{(1)}_{lm},c^{(2)}_{lm},\ldots,c^{(N_M)}_{lm}\right)^T\ .
\end{align}
The radial part of the kinetic term can be expressed as $\K \C_{lm}$, where
\begin{equation}
    \K =  \lambda^{-2}\begin{pmatrix}
 -2& \frac{2}{1} & 0 & 0 & 0& \cdots\\
 \frac{1}{2} & -2& \frac{3}{2} & 0& 0& \cdots \\
 0 & \frac{2}{3}  & -2 & \frac{4}{3} & 0& \cdots\\
 0 & 0 & \frac{3}{4} &-2& \frac{5}{4}& \cdots\\
 \vdots & \vdots & \vdots & \vdots &\vdots&\ddots
\end{pmatrix}.
\end{equation}

\section{Numerical results for various orbitals}\label{appC}

\begin{figure}[h!]
\centerline{\includegraphics[width=1.1\textwidth]{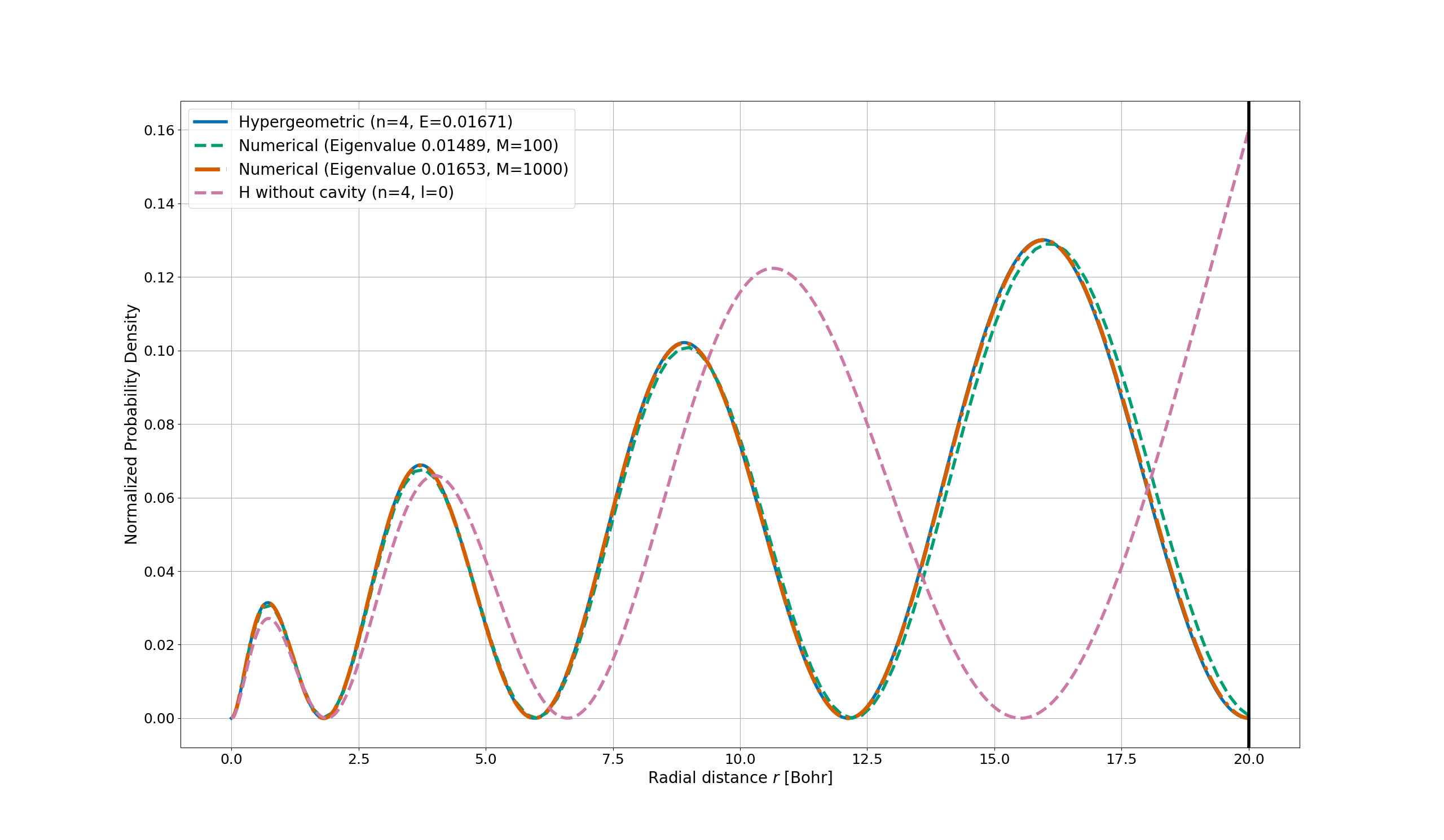}}
    \caption{Comparison of the radial part of our solution using the matrix defined  \eqref{Schrodinger} and the ordinary-space solution for confined hydrogen atom from reference \cite{aquino2007highly} for $r_0 = 20$ and \textbf{$4$S} orbital. We have also included the unconfined solution to show the effect of the barrier.}
\label{fig2}
\end{figure}

\begin{figure}[h!]
\centerline{\includegraphics[width=1.1\textwidth]{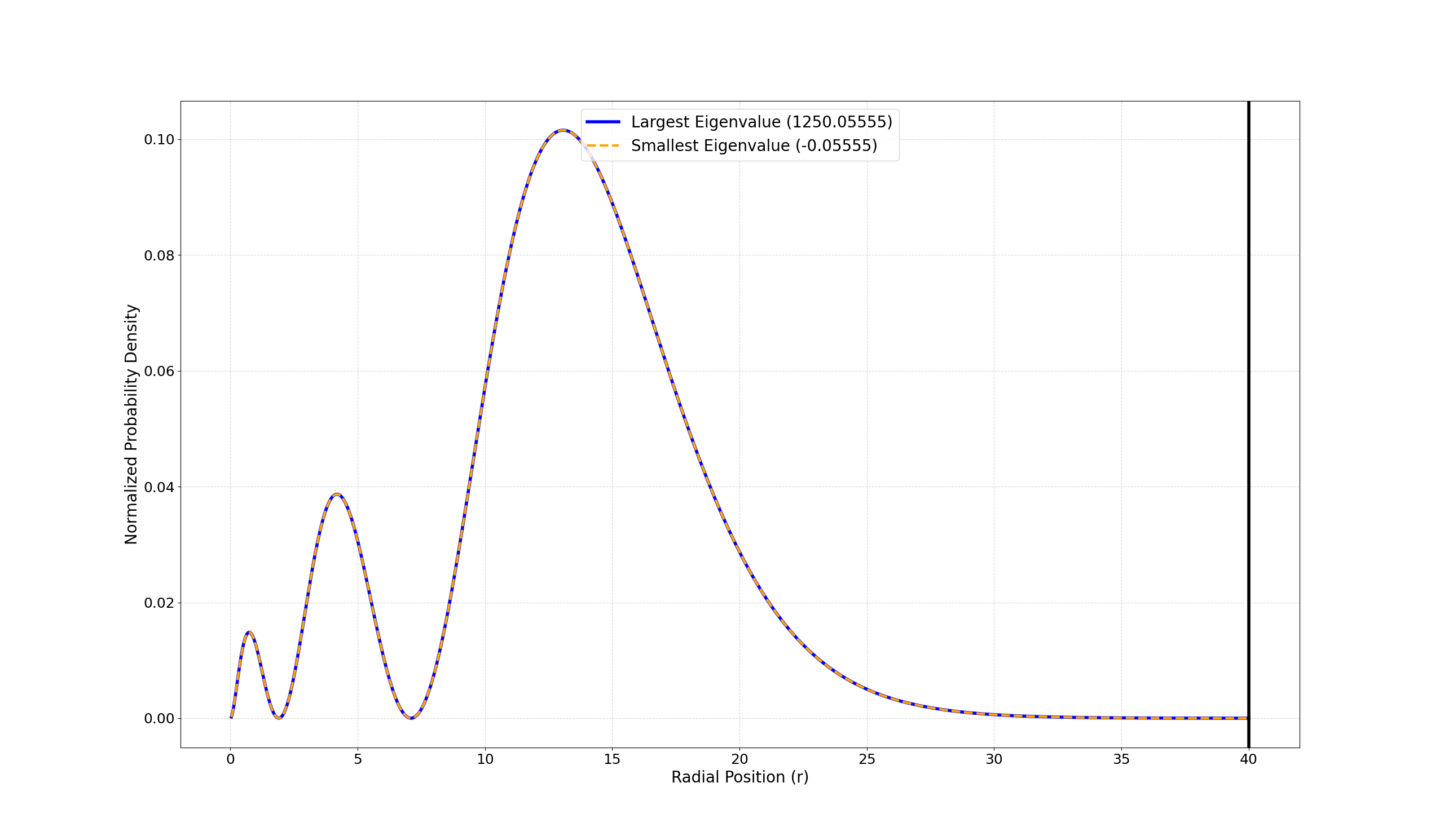}}
    \caption{Comparison of the eigenvectors for the third smallest eigenvalue in the case of attractive potential and the third largest eigenvalue for the repulsive potential; these states have dual energy according to \eqref{dual energy}. These values were obtained numerically for $M = 1000$, $r_0 = 40$ and $ n = 3$.}
\label{fig:feynman}
\label{spurious_states}
\end{figure}

The following tables \ref{tabA1}, \ref{tabA2}, \ref{tabA3}, \ref{tabA4}, \ref{tabA5} summarize the numerical results for various orbitals with the comparison with the results obtained in \cite{aquino2007highly}. After the tables, two other plots follow. Figure \ref{fig2} shows a similar situation to Figure \ref{fig1}, while Figure \ref{spurious_states} shows a comparison of two mirroring states.

\begin{table}[h!]
\resizebox{1\textwidth}{!}{
\begin{tabular}{|c|c|c|c|c|c|}
\hline
$r_0 \backslash M$    & $\infty$ as a reference   & 100                & 1000               & 10000              \\ \hline
0.5 & 14.747970030350280  & 14.406037740780091 & 14.713401904425357 & 14.744509454556344 \\ \hline
1   & 2.373990866103664   & 2.300565723232022  & 2.366554263053759  & 2.373246264259394  \\ \hline
3   & -0.423967287733454 & -0.427225951376656 & -0.424313148630359 & -0.424002075109953 \\ \hline
10  & -0.499999263281525 & -0.498755577647694 & -0.499986776756742 & -0.499999139626354 \\ \hline
20  & -0.499999999999994 & -0.495097567963923 & -0.499950009998093 & -0.499999499991737 \\ \hline
\end{tabular}}
    \caption{The \textbf{1S} orbital.}
    \label{tabA1}
\end{table}

\begin{table}[h!]
\resizebox{1\textwidth}{!}{\begin{tabular}{|c|c|c|c|c|c|}
\hline
$r_0 \backslash M$    & $\infty$ as a reference   & 100                & 1000               & 10000              \\ \hline
0.5 & 72.672039190463577 & 71.154357099658682 & 72.520341020495181 & 72.656870688127995 \\ \hline
1   & 16.570256093469736 & 16.206670008470599 & 16.533894253548414 & 16.566620019574611 \\ \hline
3   & 1.111684737436364  & 1.078613638687640  & 1.108361257317863  & 1.111352239065044  \\ \hline
10  & -0.112806210295841 & -0.113415153701996 & -0.112878188197422 & -0.112813520180460 \\ \hline
20  & -0.124987114312918 & -0.124677720985899 & -0.124984183578311 & -0.124987102906836 \\ \hline
\end{tabular}}
    \caption{The \textbf{2S} orbital.}
    \label{tabA2}
\end{table}

\begin{table}[h!]
\resizebox{1\textwidth}{!}{\begin{tabular}{|c|c|c|c|c|c|}
\hline
$r_0 \backslash M$ & $\infty$ as a reference   & 100                 & 1000                & 10000               \\ \hline
0.5 & 170.585164188274124 & 167.030984550158053 & 170.236138441457854 & 170.550329452640710 \\ \hline
1   & 40.863124601026355  & 39.992810321249898  & 40.777635017998662  & 40.854592014810407  \\ \hline
3   & 3.7349581962180043  & 3.646538108952592   & 3.726254698177943   & 3.734089328885279   \\ \hline
10  & 0.091422322407658   & 0.086434502160278   & 0.090917058740651   & 0.091371734322383   \\ \hline
20  & -0.049918047596877  & -0.050198299795152  & -0.049954888625329  & -0.049921821235481  \\ \hline
\end{tabular}}
    \caption{The \textbf{3S} orbital. }
    \label{tabA3}
\end{table}

\begin{table}[h!]
\resizebox{1\textwidth}{!}{\begin{tabular}{|c|c|c|c|c|c|}
\hline
 $r_0 \backslash M$   & $\infty$ as a reference   & 100                & 1000               & 10000              \\ \hline
0.5 & 36.658875880189399 & 35.160313617726310 & 36.505181049931707 & 36.643467765541793 \\ \hline
1   & 8.223138316160864  & 7.866678332336676  & 8.186560245882733  & 8.219471198758979  \\ \hline
3   & 0.481250312526643  & 0.449669060926531  & 0.478000341365532  & 0.480924423219991  \\ \hline
10  & -0.118859544853860 & -0.119527885630338 & -0.118934647870327 & -0.118867073891238 \\ \hline
20  & -0.124994606647078 & -0.124692831261815 & -0.124995259742673 & -0.124994633404707 \\ \hline
\end{tabular}}
    \caption{The \textbf{2P} orbital.}
    \label{tabA4}
\end{table}

\begin{table}[h!]
\resizebox{1\textwidth}{!}{\begin{tabular}{|c|c|c|c|c|c|}
\hline
 $r_0  \backslash M$   & $\infty$ as a reference    & 100                 & 1000                & 10000               \\ \hline
0.5 & 114.643552519280743 & 110.066581218096744 & 114.176502200852127 & 114.596758587994742 \\ \hline
1   & 27.473995302536328  & 26.353087093970970  & 27.359578961168989  & 27.462531765720762  \\ \hline
3   & 2.516209047333940   & 2.402055749389498   & 2.504538945119179   & 2.515039746672961   \\ \hline
10  & 0.049190760586633   & 0.042350996787297   & 0.048479625105546   & 0.049119499215287   \\ \hline
20  & -0.051611419761099  & -0.051991640223535  & -0.051665472446511  & -0.051616821787506  \\ \hline
\end{tabular}}
    \caption{The \textbf{3P} orbital.}
    \label{tabA5}
\end{table}

\end{document}